\documentstyle[12pt,psfig]{article}
\def\bold#1{\setbox0=\hbox{$#1$}%
      \kern-.025em\copy0\kern-\wd0
      \kern.05em\copy0\kern-\wd0
      \kern-.025em\raise.0433em\box0 }
\def\eea{\end{eqnarray}}
\def\bea{\begin{eqnarray}}
\def\eeas{\end{eqnarray*}}
\def\beas{\begin{eqnarray*}}
\def\ee{\end{equation}}
\def\be{\begin{equation}}
\def\bdm{\begin{displaymath}}
\def\edm{\end{displaymath}}

\begin{document}
%%%%%%%%%%%%%%%%%%%%%%%%%%%%%%%%%%%%%%%%%%%%%%%%%%%%%%%%%%%%%

\begin{titlepage}
\begin{center}
\vspace*{2.0cm}
{\Large\bf On the classical dynamics of billiards on\\
the sphere }
\vskip 1.5cm

{M.E.Spina and M.Saraceno}
\vskip .2cm
{\it Department of Physics, \\ Comisi\'on Nacional de Energ\'{\i}a At\'omica,
          Av. Libertador 8250, \\
(1429) Buenos Aires, Argentina. }
\vskip 2.cm
March 1999 \\
\vskip 2.cm
{\bf ABSTRACT}\\
\begin{quotation}

We study the classical motion in bidimensional polygonal billiards on the sphere.
In particular we investigate the dynamics in tiling and generic rational and irrational
equilateral triangles. Unlike the plane or the negative curvature cases we obtain 
a complex but regular dynamics.

\end{quotation}

\end{center}
\vspace{1cm}
%{\it PACS}:  \\
%{\it Keywords}: 
 
\end{titlepage}

%%%%%%%%%%%%%%%%%%%%%%%%%%%%%%%%%%%%%%%%%%%%%%%%%%%%%%%%%%%%%%
\section{INTRODUCTION}

In this work we consider the classical motion of a point particle 
inside a two-dimensional polygonal billiard on a surface with
constant positive curvature. Flat billiards and billiards on a surface with
negative curvature have been extensively studied.
It is well known  that the dynamics in flat polygonal billiards 
depends on whether the inner angles of the polygon are rational 
multiples of $ \pi $  \cite{bo,gu8,gu9}. If this is the case these systems are referred 
to as 'pseudointegrable' since they possess two constants of motion and the
flow is restricted to a two-dimensional invariant surface. 
If at least one of the vertex angles is an irrational multiple
of $ \pi $  the polygon is believed to possess ergodicity \cite{casa}.
On the other hand 
the interest in studying polygonal billiards on a surface with negative curvature is that the 
classical motion is simple and as chaotic as possible since the flow on these
surfaces is hyperbolic \cite{voros}.

\noindent By investigating the dynamics on a spherical surface we explore the other limit: 
instead of having more chaoticity, more focusing is expected with respect to the planar
case as a consequence of the positive curvature.
The motivation for studying 
these systems is to see how this focusing mechanism 
together with the compactness of the surface affects the classical motion. 
In particular we will investigate the condition for 
integrability and stress the importance of tiling billiards.
In the general case we will explore numerically the phase space portrait of curved polygonal 
billiards (i.e., periodic orbits, invariant surfaces, etc),  
and try to explain these numerical observations in a rather intuitive way.\\

\noindent The paper is organized as follows. In section 2 we present the formalism. In section 3 
we briefly discuss the case of billiards enclosed by meridians and parallels and then concentrate
on the study of polygonal billiards, i.e. billiards whose boundary consists entirely of
arcs of
geodesics. In particular we will focus on equilateral triangles. We will first study tiling
triangles in subsection 3.1 and then generic triangles in subsection 3.2.
 For this we use the entry-exit map and finally 
make a comparison of small curved triangles with triangles on the plane.
Conclusions are presented in section 4.

\section{THE MODEL}

The motion on the sphere will be described in terms of the spherical coordinates: 
the polar angle $\alpha$ and the azimuthal angle $\beta$. 
The coordinate curves $\alpha=const$ and $\beta=const$ form an 
orthogonal net. The line element for a sphere of radius $ R $ has the usual form:

\begin{equation}
ds^2= R^2  d\alpha^2+ R^2 \sin^2 \alpha d\beta ^2
\end{equation}

\noindent and the curvature is $1/R^2$.

\noindent The geodesics are the great circles of the sphere and can be viewed as its
intersection with a plane passing through the origin. They are labeled
by the coordinates $ (\theta^G,\phi^G) $ of the unit vector normal to this plane and  
their equation reads:

\begin{equation}
\tan \alpha= -{ \cot \theta^G \over \cos(\beta-\phi^G)}
\end{equation}

\noindent Note that $ \alpha$ and $ \beta$ are coordinates on the physical sphere 
where the billiard lies 
while $ \theta^G $ and $ \phi^G $ denote a point on what we will call the dual 
sphere. This is illustrated in Fig.1.
In the problem we are considering the particle moves freely along
an oriented geodesic labeled by $ (\theta^G,\phi^G) $ until it suffers a specular 
reflection at the boundary and jumps on a different geodesic.
Point $ (\theta^G,\phi^G)_i $ maps into a new point $ (\theta'^G,\phi'^G)_{i+1} $ and
a given trajectory can be specified uniquely by an infinite succession of pairs
$ (... (\theta^G,\phi^G)_j...)$ on the dual sphere 
labeling the successive geodesics on which the motion takes place. In this way 
the motion is completely described by a map acting in discrete time on a bounded
domain on the surface of the dual sphere. The area element $ dA= \sin \theta d\theta
 d\phi $ is preserved by this bounce map $ T $.
 Thus $ \cos \theta^G $ and $\phi^G$
are canonical coordinates on the dual sphere. They provide a convenient reduced
description of the classical motion alternative to the Birkhoff coordinates.

\section{POLYGONAL BILLIARDS}

The simplest systems on the sphere one can think of are
convex billiards whose boundary consists of meridians ($ \beta=const$) and 
parallels ($\alpha=const$). These are not polygons since parallels (except for the
 equator) are not geodesics.
It is easy to see that in reflections on meridians and parallels, the
 quantity $\sin \theta^G$ is  conserved . Therefore, the dynamics 
for these billiards is integrable.
Depending on the initial conditions we get periodic and non periodic orbits, giving rise
to rational and irrational tori in phase space.

\noindent We now concentrate on polygonal billiards, i.e. systems enclosed 
by arcs of geodesics. Note that in this case not only the trajectories but also 
the boundaries can be specified by points on the dual
sphere. Therefore 
the problem can be viewed as a mapping of unit vectors by a matrix also 
expressed in terms of coordinates on the dual sphere.
This would not be possible in the general case of a billiard with arbitrary enclosure.

\noindent For simplicity we will restrict our discussion to the case of equilateral triangles.
While flat equilateral triangles are integrable, spherical equilateral triangles 
 present a rich variety of possibilities depending on their size. This is due to the
existence on the sphere of a definite
relation between the size and the shape of the triangle.
The inner angle $\omega$ can take any value in the interval $ [{\pi \over 3}, \pi]$, 
and the area  (that can be related to the total curvature) is:
$ A = R^2 (3 \omega - \pi) $.

\noindent For a triangle centered on the north pole 
the three sides will be specified by the vectors
$ (\theta^B, \phi^B_i) $ which label the intersecting
geodesics. Here, $ \phi^B_i=(i-1){2 \pi \over 3}$ with $ i=1,2,3 $, and  
the angle $ \theta^B $ is related to the inner angle $\omega$ through:

\begin{equation}
\sin \theta^B = {\cos {\omega \over 2} \over \cos {\pi \over 6}}
\end{equation}

\noindent Vertex $V_i$ defined as the intersection of geodesics $ (i-1) $ and $ i $ will
be located at $(\alpha^V, \beta^V_i)$ , where $(\alpha^V, \beta^V_i)$ denote coordinates 
on the physical sphere given by:

\be
\tan \alpha^V={2 \over \tan \theta^B}
\ee 
and 

\be 
\beta^V_i= \phi^B_i+{2 \pi \over 3}
\ee

\noindent It will be useful in the following to 
 introduce for each vertex $ V_i $ the curve $ C_i $, locus of
 the points $ (\theta, \phi) $ specifying all the geodesics passing by $ V_i $.
 Its equation reads:

\be
\tan \theta_i= -{ \cot \alpha^V \over \cos(\phi_i-\beta^V_i)}
\label{curvei}
\ee

\noindent It follows from Eq.(\ref{curvei}) that curve $ C_i $ is itself a geodesic on the dual 
sphere labeled 
by $ (\alpha^V,\beta^V_i) $. Note that this is not true in general but rather a consequence
of the metric of our problem that will play an important role when understanding the
structure of phase space.
Once we have defined the discontinuity curves $ C_i $
it is easy to see that a point $ (\theta,\phi) $ corresponds to a geodesic 
entering (exiting) side $ i $ if:
$ \theta_i (\phi) \leq \theta \leq \theta_{i+1} (\phi) $ 
( $ \theta_{i+1} (\phi) \leq \theta \leq \theta_i (\phi) $ ) .

\noindent The area preserving mapping $ T $ can be written explicitely as an orthogonal
matrix  
acting on the unit vector 
$ \left( \begin{array}{c} \sin \theta \cos \phi\\
                         \sin \theta \sin \phi  \\
                          \cos \theta
       \end{array}
\right) $ as:

\be
T = T_i = \left( \begin{array}{ccc} \cos \phi^B_i & -\sin \phi^B_i & 0 \\
                          \sin \phi^B_i & \cos \phi^B_i & 0 \\
                          0 & 0 & 1
       \end{array}
\right)
\cdot
\left( \begin{array}{ccc} -\cos 2 \theta^B & 0 & \sin 2 \theta^B \\
                          0 & -1 & 0 \\
                          \sin 2 \theta^B & 0 & \cos 2 \theta^B
       \end{array}
\right)
\cdot
\left( \begin{array}{ccc} \cos \phi^B_i & \sin \phi^B_i & 0 \\
                          -\sin \phi^B_i & \cos \phi^B_i & 0 \\
                          0 & 0 & 1
       \end{array}
\right)
\label{map}
\ee

\noindent for $\theta_i (\phi) \leq \theta \leq \theta_{i+1} (\phi)$, that is for 
a geodesic entering side $i$.

\noindent Since the problem is symmetric under rotations in $ {2 \pi \over 3}$ around 
the center of the triangles it will be useful to define 
the operators $ T^+ $ and $T^-$ as :

\be
T^{\pm}= \left( \begin{array}{ccc} \cos{2 \pi \over 3} \ & {\pm}\sin{2 \pi \over 3} & 0 \\
                          {\mp}\sin{2 \pi \over 3} & \cos{2 \pi \over 3} & 0 \\
                          0 & 0 & 1
       \end{array}
\right)
\cdot
\left( \begin{array}{ccc} -\cos 2 \theta^B & 0 & \sin 2 \theta^B \\
                          0 & -1 & 0 \\
                          \sin 2 \theta^B & 0 & \cos 2 \theta^B
       \end{array}
\right)
\ee

\noindent With this definition according to which, for example, $ T_3 T_2 T_1 = T^+ T^+ T^+ $
 and $ T_2 T_1 = T^+ T^- $, we indicate if the
particle exiting from a given side hits the next or the previous side in increasing order
 with $i$.

\noindent The trajectories in the triangle will be classified according to an infinite
symbol sequence obtained by listing the  sides successively hit.
 The code alphabet will consist of $ + , - $ signs. This classification, as 
we will see, is not one to one in the sense that even infinite length sequences do not 
distinguish single trajectories uniquely.

\subsection{TILING TRIANGLES}

We now study the case of equilateral triangles that tile the sphere under the
reflection rule. Since the rotation group has a finite number of discrete subgroups 
there are only a few ways of tessellating the sphere. Tiling triangles are such that their
vertices coincide with those of a face of a regular polyhedron. The three possible 
cases are the tetrahedron with $ \omega = {2 \pi \over 3}$, the octahedron with
$ \omega = {\pi \over 2}$ and the icosahedron with $ \omega = {2 \pi \over 5}$.
We will see that tiling triangles constitute a very particular class of billiards 
that not only are integrable but for which only periodic orbits are present.

\noindent To study the motion in these three particular triangles we follow the procedure 
presented in \cite{bo} : every time an edge is hit, instead
of reflecting the incident geodesic, we reflect the billiard across the edge and follow
the same geodesic into the replica. That is, the motion is viewed as a unique geodesic
 entering and exiting copies of
the original billiard. Since the surface is compact and the billiard is tiling
it is clear that only periodic 
orbits exist. In order to determine their periodicity $ n_p $ we have to fold back, 
for each copy, the corresponding segment of geodesic into the original billiard.
When, after a certain
number of these operations, the image coincides with the original 
the trajectory closes. To do this we have to consider the 
symmetry group of the corresponding polyhedron and the possible circuits on it, that is,
which faces are visited by the geodesic.

\subparagraph{ Triangle with $\omega= {2 \pi \over 3} $}.  

\noindent Two circuits are possible on the projected tetrahedron: one visits 3 faces, 
corresponding to orbits of type  $[...  + -  ...]$ in the triangular billiard,
 and the other 4, corresponding to orbits  $[...  + + + ...]$ . Both
circuits are indicated in Fig.2, where the notation for the tetrahedron
is defined. Sides are indicated with $ a,b,c...$, while with $\sigma_a, \sigma_b, \sigma_c...$ 
we denote reflections across these sides. Face $I$ corresponds to the original billiard.

The orbits of the first type correspond to following a geodesic passing
through faces : $ I , \sigma_b I , \sigma_c \sigma_b I , \sigma_a \sigma_c \sigma_b I ,
...$.
The transformation $ \sigma_j \sigma_i $ is a rotation $ C_3^2 $
through $ {4 \pi \over 3}$
about the line of intersection of the reflection planes.
Thus the product of 6 reflections, corresponding to a rotation in $ 4 \pi $, brings
back the segment of geodesic to its original position and orientation. Therefore,
general periodic orbits of type $[...  + - ...]$ have periodicity 
$ n_p = 6 $.

There are three particular orbits of lower periodicity $ n_p = 2 $.
 These are the ones invariant
under the rotation around three of the four $ C_3 $ axes of the tethahedron, that is, the 
geodesics
resulting from the intersection with the sphere of the planes orthogonal to these axes. 

\noindent The orbits of type $[...  + + + ...]$ correspond to a geodesic passing through 
faces: $ I , \sigma_b I , \sigma_c \sigma_b I$ , $\sigma_f \sigma_c \sigma_b I ,
\sigma_a \sigma_f \sigma_c \sigma_b I ,...$.
It can be seen that
the product of three reflections  $ \sigma_k \sigma_j \sigma_i $ is a rotary-reflection
$ S_4 $ about a one of the binary axis joining the centers of two opposite sides of the
tetrahedron, as shown in Fig.2. Since $ (S_4)^4 = E $, 12 reflections are needed in order
to recover the initial segment of geodesic in the original billiard and the periodicity
of the orbits will be $ n_p =12 $.
There is, finally, an orbit with periodicity $ n_p =3 $ that joins the centers of the tree 
sides of the triangle.

\subparagraph{Triangle with $ \omega= { \pi \over 2} $}.  

Due to the geometry of the octahedron, all the trajectories in this triangle
are of the type $[...  + + + ...]$. A product of three reflections is equivalent
to an inversion, transforming $ \theta^G, \phi $ in $ \pi - \theta, \phi + \pi $.
Since any geodesic is invariant under this operation, the periodicity for all
the orbits will be $ n_p = 3 $.

\subparagraph{Triangle with $ \omega= {2 \pi \over 5} $}. 

Following the same procedure we find in this case orbits of type $[...  + + + ...]$ 
which have in  general periodicity  $ n_p = 15 $ (and one with  $n_p = 3 $ that joins 
the centers
of the three sides of the triangle) and orbits of the type  $[... + + - -...]$
with  $ n_p = 12 $ (and three orbits with lower periodicity $ n_p = 4 $).

\noindent One can also think of triangles that are not tiling in the strict sense but, 
by successive reflections on their sides, cover the sphere more than once.
This is the case of a triangle with $\omega={4 \pi \over 5}$ that covers 
the spherical surface twice.
 It has periodic orbits of type $ [... + - ...] $ with periodicity $n_p=10$,
of type $ [...+++...] $ with periodicity $n_p=15$ and of type $ [...+ + - + + - + + - ...] $ 
with periodicity $n_p=9$.

\subsection{GENERIC TRIANGLES}

Let us now investigate the case of a generic triangle with an arbitrary $\theta^B$.
Following \cite{ullmo}
we will divide the available phase space in the $ (\theta,\phi) $- plane 
 in entry (or exit) domains of the three sides. The enter (exit) domain is defined as the set 
of points associated with the oriented geodesics that enter
(or exit) a given side. 
In order to do this partition, we use curves $C_i$, defined in Eq.(\ref{curvei}).

\noindent In FIG.3a we show an example of an entry-exit map.
The intersections of curves $C_i$ and $C_{i-1}$ correspond to the points associated to
side $ i-1 $ with the two possible orientations.
Each point in the available phase space belongs to one entry and one exit domain.
Each exit domain is intersected by two entry domains.

\noindent After $n$ iterations of the map according to Eq.(\ref{map}) the phase space gets 
partitioned into domains which are enclosed by geodesics, since the images by reflection of the 
separating curves $ C_i$ are also geodesics.
Each domain can be labeled by a sequence of $n$ symbols denoting the $n$
sides successively hit by the trajectory.  

\noindent The only general rule limiting the possible sequences is that a side cannot be hit 
twice consecutively. But it is clear that for each particular triangle the existence or absence
of a region associated with a given sequence is determined by the geometry, in this case,
the inner angle $ \omega $. 
For example the domain corresponding to an infinite periodic sequence of type
 $ [... +- ...]$ exists only in triangles with angle $\omega> {\pi \over 2}$. 

\noindent FIG.3b shows the partition of phase space after two iterations of the map. 

\noindent As long as the number of iterations remains finite the different allowed domains
are bounded by arcs of geodesics and therefore polygons of increasing complexity.
\noindent When a given sequence of symbols is repeated periodically an infinite number
of times the resulting set of trajectories is a chain of elliptic islands bounded by a
smooth curve which is the limit of the above mentioned polygons. Short repeated sequences
lead to large and fairly regular areas, while long sequences lead to complex island chains.
In Fig.3c shows the domains corresponding to the repetition of some short codes.

\noindent Before studying the structure of these islands in more detail, let us have a look at 
the entry-exit maps for the special case of the tiling triangles treated in the previous
 section. The common feature to these particular triangles is that the corresponding separating 
geodesics $C_i$ are reflected into themselves after a few iterations. Therefore
phase space gets partitionned into a few  polygonal (in the sense that they are enclosed
by a finite number of segments of geodesics) domains corresponding to different periodic 
codes. Inside each domain, all orbits 
are periodic with the same periodicity and we recover the result obtained in the previous 
section. This is illustrated in FIG.4. 
Other rational but non-tiling
triangles may have domains in which all trajectories are periodic, coexisting
with the families of elliptic islands. These triangles with periodic domains can be found by 
requiring the 
angle $\theta^B$ to be such that successive applications of the bouncing matrix corresponding
to a chosen sequence gives the unit matrix.
For example requiring that $ ( T^+ T^+ T^+ )^n = I $ gives
\be
\theta^B= \arccos { \cos {k \pi \over n} \over \cos {\pi \over 6}}
\label{con1}
\ee

while $( T^+ T^- )^n= I $ fixes

\be
\theta^B= \arcsin { \cos {2 k \pi \over {3 n}} \over \cos {\pi \over 6}}.
\label{con2}
\ee

\noindent where $k$ and $n$ are integers.

\noindent Thus, if a triangle defined by $ \theta^B $ and its complement, i.e., the one 
defined by $ {\pi \over 2} - \theta^B $  have inner angles $ \omega_1 $
and  $ \omega_2 $ which are both rational,
 angle $ \theta^B $ satisfies simultaneously Eq.(\ref{con1}) and Eq.(\ref{con2})
and the  pair of rational angles $(\omega_1 ,\omega_2)$ fulfill: 
\be
\cos \omega_1 + \cos \omega_2 = -{1 \over 2}
\label{con3}
\ee

\noindent We found numerically two pairs of rational angles satisfying Eq.(\ref{con3}):
 $(\omega_1 = {\pi \over 2}, \omega_2 ={2 \pi \over 3})$ and 
$(\omega_1 = {2 \pi \over 5}, \omega_2 ={4 \pi \over 5})$, corresponding to the triangles
 analyzed in the previous section.

\noindent Summarizing 
we see that the entry-exit map is a useful tool to determine whether, according 
to the geometry (in this simple case the inner angle $\omega$), a domain corresponding
to a given code exists or not in a given triangle.
For this it is sufficient to reflect the segments of the separating geodesics
$ C_i $ enclosing the initial domain according to the chosen
sequence.
The image domain after infinite iterations can be either empty or finite, in contrast with 
the hyperbolic case where non empty domains associated to a code are reduced to a single
point. Finite domains are in general bounded by smooth curves, 
resulting from the intersection of an infinite 
number of geodesics. In rational triangles some particular codes are associated to polygonal 
domains enclosed by a finite number of geodesics.

\noindent In order to understand the detailed structure of phase space in the generic case 
we now follow individual trajectories. 
A phase space plot is shown in Fig.5 for different initial conditions. 
It reminds very much the stable regime of a sawtooth map \cite{laks}. 
At the center of each island, belonging to a set of $n_p$ islands,
sits a periodic orbit of period $n_p$   
surrounded by a family of nested invariant curves. These correspond to
open orbits having an infinite but periodic sequence of period
$n_p$ . That is, a given code does not specify a 
single trajectory, as in the hyperbolic case, but an entire family of orbits.
The size of the islands decreases as $n_p$ increases  
and phase space takes a fractal structure.

\noindent The existence of these islands is a consequence of the focusing mechanism on the
spherical surface. The fact that two geodesics on the sphere intersecting at
$ \theta_{int}, \phi_{int} $ cross again at $ \pi - \theta_{int}, \phi_{int} + \pi $ 
reflects on the structure of the orbits.
To illustrate this we consider a periodic orbit with initial conditions
$ ( s^{p.o.}, p_{\parallel}^{p.o.} )$
 expressed in Birkhoff coordinates and an open trajectory close to it.
If the initial 
conditions are close enough to the ones of the periodic orbit the particle on the open 
trajectory will follow the 
sequence of the periodic orbit and, after a period, end at a point $ s $ on the boundary that
is shifted with respect to the point $ s^{p.o.} $.
 The position $ s $ as a function of 
the number of period iterations is plotted in Fig.6 for a given orbit in a curved and a flat
triangle of equal area
. In the flat billiard the shift $ s - s^{p.o.} $ increases linearly, in
such a way that, after a certain number of iterations, the vertex will eventually be reached 
 and the original sequence broken. In the spherical case the shift is an oscillatory function
of the number of iterations. If the amplitude of oscillation is small enough the corner is
never reached: the pendulating motion lasts forever and the particle keeps
repeating the original symbolic sequence but never retracing itself exactly.
This is what happens in the islands of Fig.5. As long as the concentric curves do not
become tangent to curves $C_i$, that act as separatrices, the open trajectories have the
same code as the central periodic orbit. When the outer curve gets tangent to $C_i$, the 
original sequence is broken and a new and more complex sequence appears, corresponding to
a new chain of islands.

\noindent The corresponding transformation is a product of a string of orthogonal matrices
$ T_i $. The resulting $ 3 \times 3 $ matrix is itself orthogonal and therefore has an 
eigenvalue $ 1 $, corresponding to preservation of the norm, and a pair of complex
conjugate phases $ e^{{\pm}i \Omega} $. Thus the motion inside each chain of islands is
labeled by a code and is everywhere elliptic with the same rotation angle $\Omega$.
This is clearly displayed ing Fig.5 and we have verified it numerically for several
 short sequences.

\noindent An alternative way to analyze the phase space structure of triangular billiards 
on the sphere is to go to the 
limit of small triangles where the total curvature can be seen as a perturbation from the 
integrable case of flat equilateral triangles. 
This limit corresponds to take $\omega \rightarrow {\pi \over 3}$, (that is 
$\theta^B = {\pi \over 2}-\epsilon$). This restricts the available space phase to 
geodesics with $ \theta^G $ close to ${\pi \over 2}$.
For the segments inside the billiard, that is $\beta$ in a small
interval around $\phi^G + \pi$, the equation of the geodesics reads:

\be
\alpha = {\theta^0 \over \cos(\beta - \phi^0)}
\ee

\noindent where $\theta^0 = {\pi \over 2} - \theta^G = \epsilon$ and
 $\phi^0 = \phi^G + \pi$, that is, the equation 
of a straight line in polar coordinates in the plane.

\noindent The reflection across a side of the planar triangle is given by:
\be
\alpha'= -\alpha + 2 \theta_0 \cos(\beta-\phi^0) 
\ee

\be
\beta' = -\beta + 2 \phi_0
\ee

\noindent which can also be obtained by expanding Eq.(\ref{map}) to first order in $\epsilon$.
All trajectories lie on tori filling phase space. 
In Fig.7a  several rational and irrational tori are shown.
Under perturbation, these rational tori are destroyed.
 In contrast with the generic case 
for which elliptic and hyperbolic cycles of fixed points survive,
fixed points that survive perturbation in the spherical case 
are all stable. The absence of hyperbolic trajectories might be attributed to 
the fact that Lyapunov exponents
are imaginary as a consequence of the focusing mechanism due to the positive curvature.
This point needs further investigation.
 
\noindent Around each stable point of the periodic orbit $n_p$
a family of  invariant curves, corresponding to open trajectories, develops.
(that is, we get
a chain of $n_p$ islands). This is shown in Fig.7b.
As mentioned in the previous section, 
the periodic code corresponding to the central periodic orbit characterizes 
the whole family.

\section{CONCLUSIONS}
This work was a first step in the understanding of the classical dynamics in billiards
on a spherical surface. We only considered polygonal billiards and, more specifically,  
equilateral rational and irrational triangles. The main observation was the existence in
the vicinity of each periodic orbit of a set of open orbits characterized by the same 
infinite periodic code. This is a consequence of the focusing mechanism and constitutes a
substantial difference with respect to the planar and hyperbolic cases where each infinite
code is associated to a single orbit. The map has a regular but very complex structure.
The phase space is covered by chains of elliptic islands, whose multiplicity increases and
size decreases with the period. Tiling billiards constitute a very special system where only
periodic orbits are present. 
We have also studied non-equilateral triangles and the general structure of phase space is
very similar. We do expect differences if the boundaries are not geodesics: in particular
the motion is likely to develop large chaotic regions. This question together with the issues
of quantization will be addressed in the future.

\pagebreak

\pagebreak

\vskip .1cm
\centerline{ { \psfig{figure=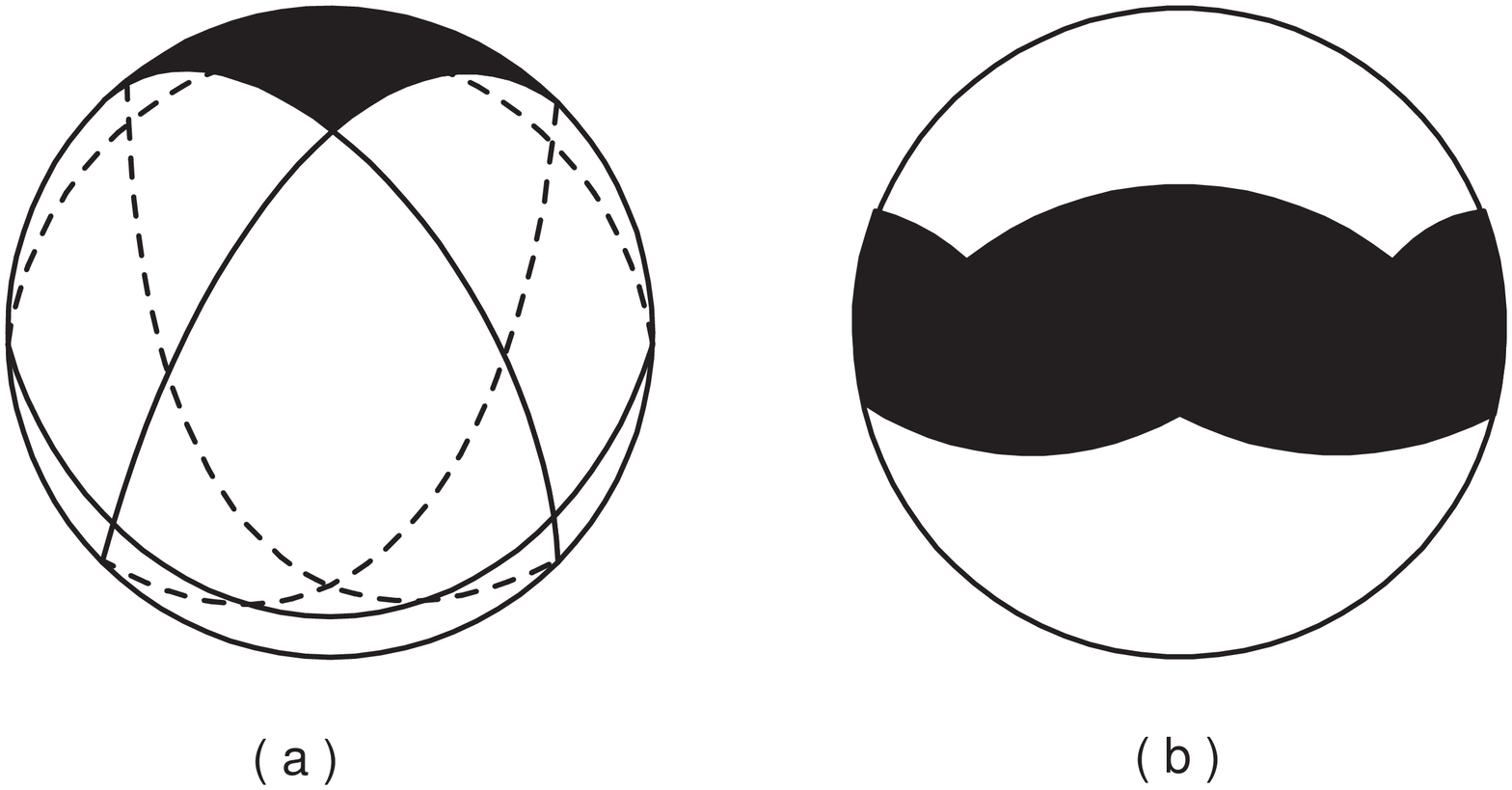,width=13.cm,height=7.cm,angle=0} } }
\vskip .5cm
\centerline{ \parbox{15cm} 
{{\normalsize {Fig.1 -- (a) shows three intersecting geodesics enclosing a triangular
billiard centered at the north pole.
In (b) the shadowed area is the corresponding phase space on the dual space
to the geodesics.} }} }

%\pagebreak

\vskip 2.cm
\centerline{ { \psfig{figure=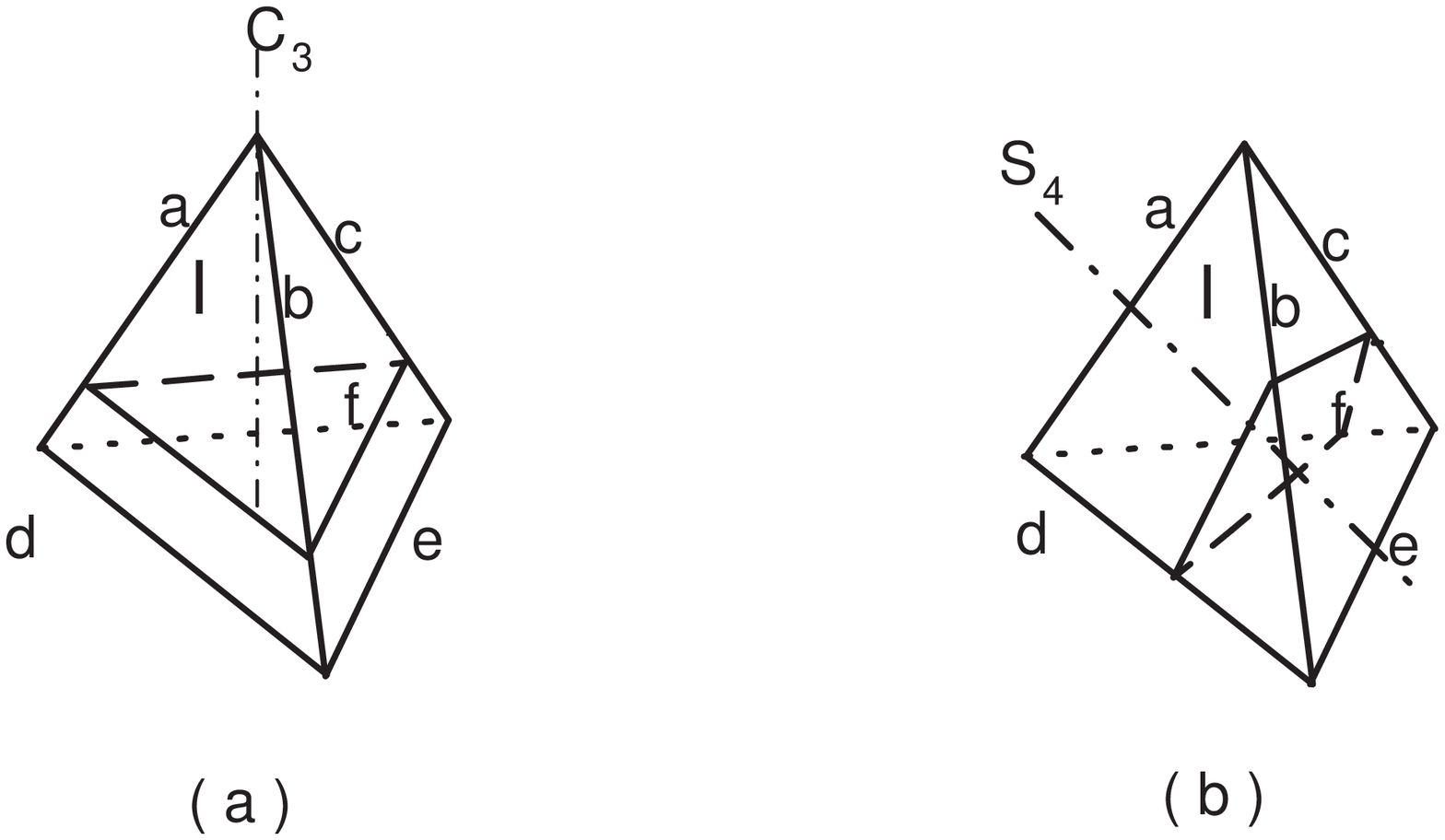,width=13.cm,height=7.cm,angle=0} } }
\vskip .5cm
\centerline{ \parbox{15cm} {{\normalsize 
{Fig.2 -- (a) shows schematically a circuit visiting 3 faces of the tetrahedron, corresponding
to an orbit of type [...+ -...] in triangle I. In (b) a circuit visiting 4 faces, associated to 
an orbit of type [...+ + +...]. } }} }

%\pagebreak

\vskip .1cm
\centerline{ { \psfig{figure=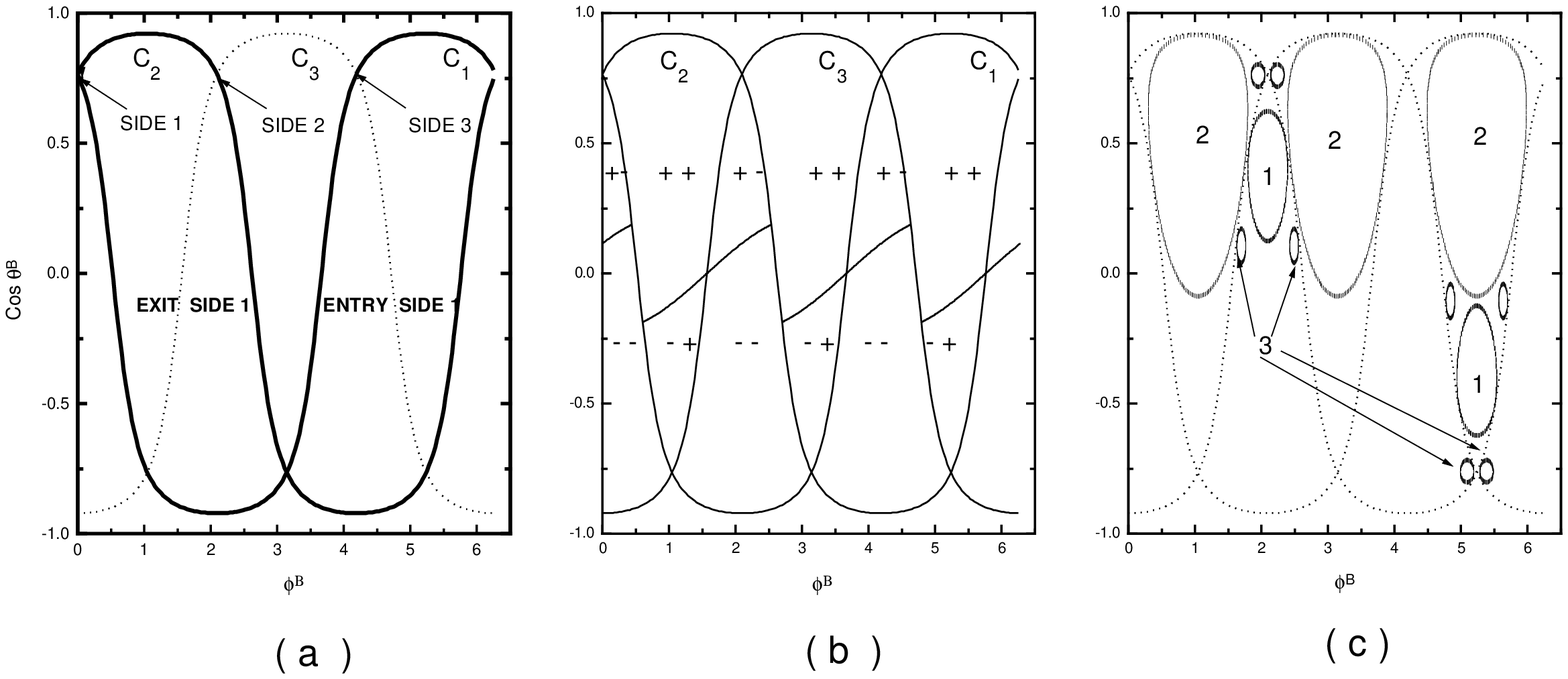,width=18.cm,height=15.cm,angle=0} } }
\vskip .5cm
\centerline{ \parbox{15cm} {{\normalsize {Fig.3 --
(a) Entry-exit map for a triangle with $\theta^B=0.7$. 
(b) Phase space after two iterations gets partitionned in domains corresponding
 to orbits of type [+ +] ([- -]) and of type [+ -] ([- +]).
(c) shows the domains corresponding to orbits [...+ -...] (1), [...+ + +...] (2) and
[... + + + + - - - -...] (3) after infinite iterations.} }} }

%\pagebreak

\vskip .1cm
\centerline{ { \psfig{figure=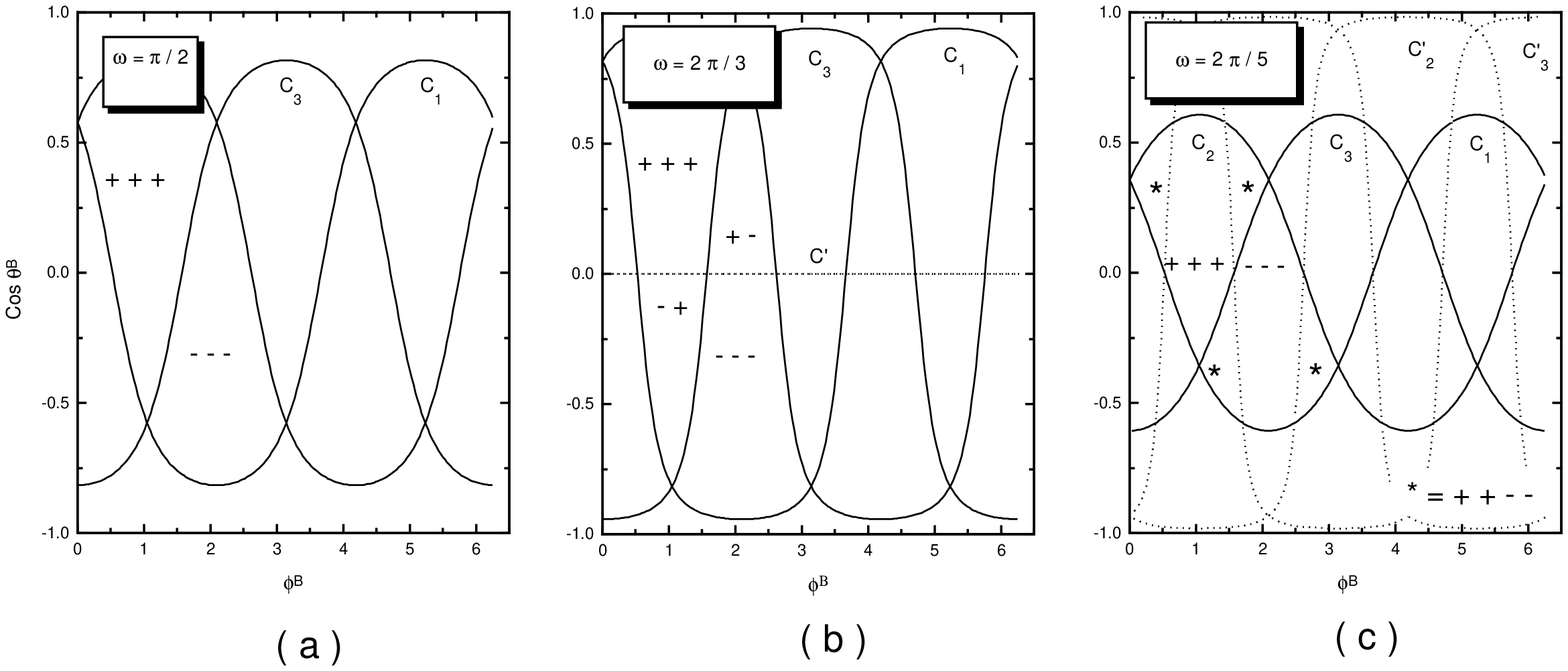,width=18.cm,height=15.cm,angle=0} } }
\vskip .5cm
\centerline{ \parbox{15cm} {{\normalsize {Fig.4 --
Phase space portrait for tiling triangles. (a) For $ \omega = {\pi \over 2 } $ 
curves $ C_i$ reflect into themselves. (b) For $ \omega = {2 \pi \over 3 } $
curves $ C_i$ reflect into $ C'$ ($\theta^B = {\pi \over 2} $). (c) For 
$ \omega = {2 \pi \over 5 } $ curves $C_i$ reflect into $C'_i$ which, in turn, reflect
again into $C_i$.} }} }

\vskip .1cm
\centerline{ { \psfig{figure=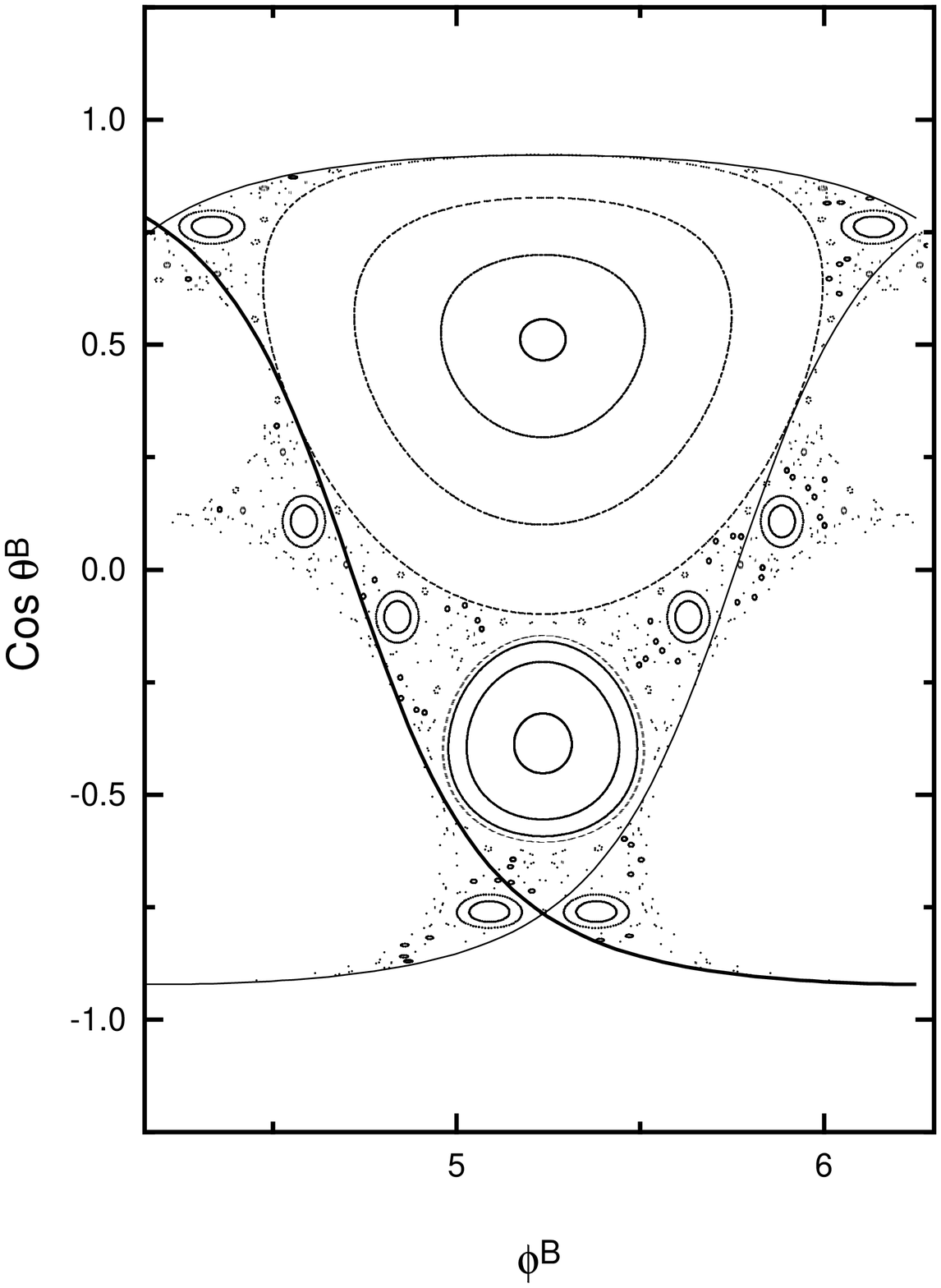,width=15.cm,height=20.cm,angle=0} } }
\vskip .5cm
\centerline{ \parbox{15cm} {{\normalsize {Fig.5 -- Phase space plot for a triangle
with $\theta^B=0.7$. 20 orbits are shown, each iterated 1000 times.} }} }

%\pagebreak

\vskip .1cm
\centerline{ { \psfig{figure=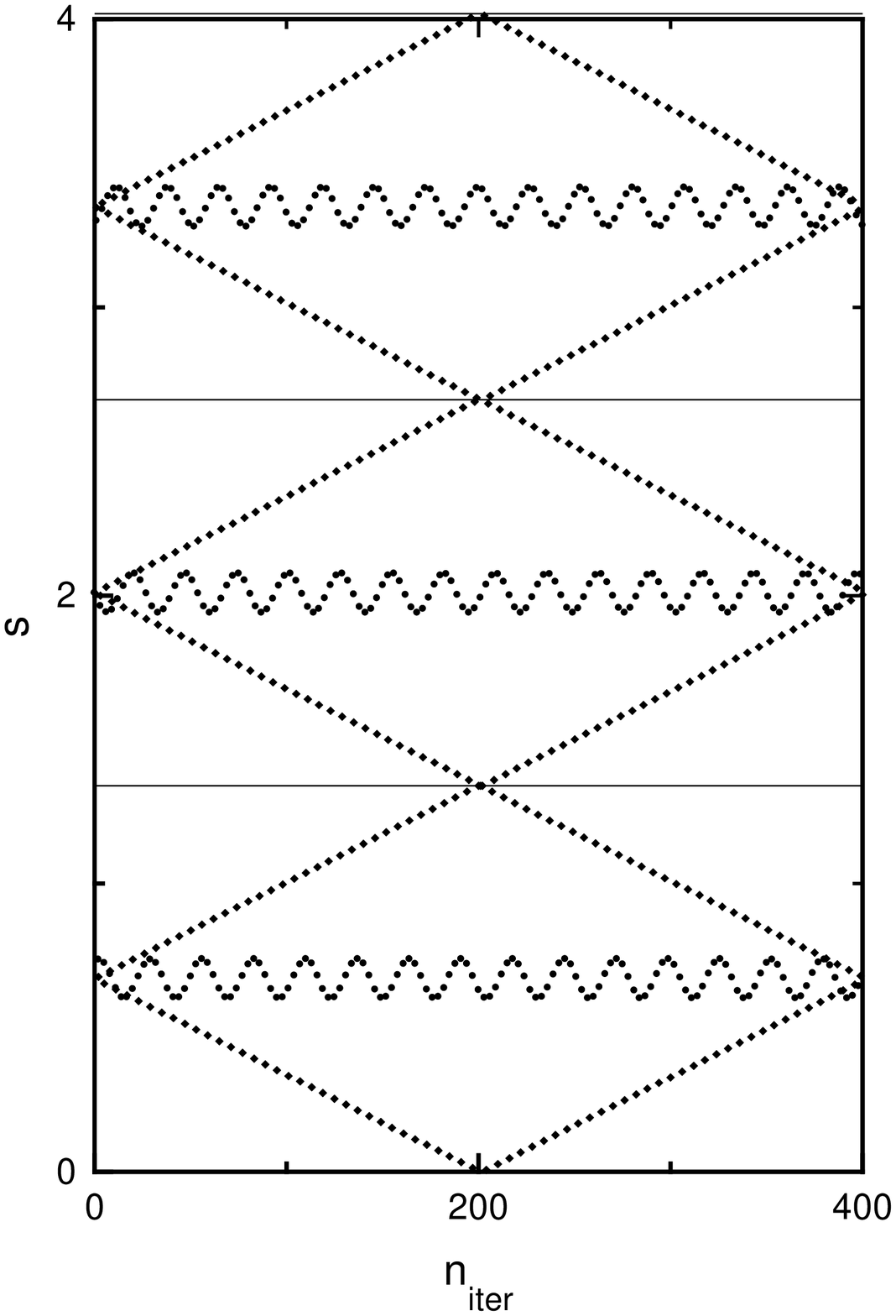,width=15.cm,height=18.cm,angle=0} } }
\vskip .5cm
\centerline{ \parbox{15cm} {{\normalsize {Fig.6 -- Position on the boundary $s$
plotted as a function of the number of iterations for an open trajectory in the 
neighborhood of a periodic orbit of period $n_p=3$, starting at $s_{p.o.}$.
Circles correspond to the curved triangle, diamonds to the planar triangle of equal area.
Horizontal lines indicate the edges.} }} }

%\pagebreak

\vskip .1cm
\centerline{ { \psfig{figure=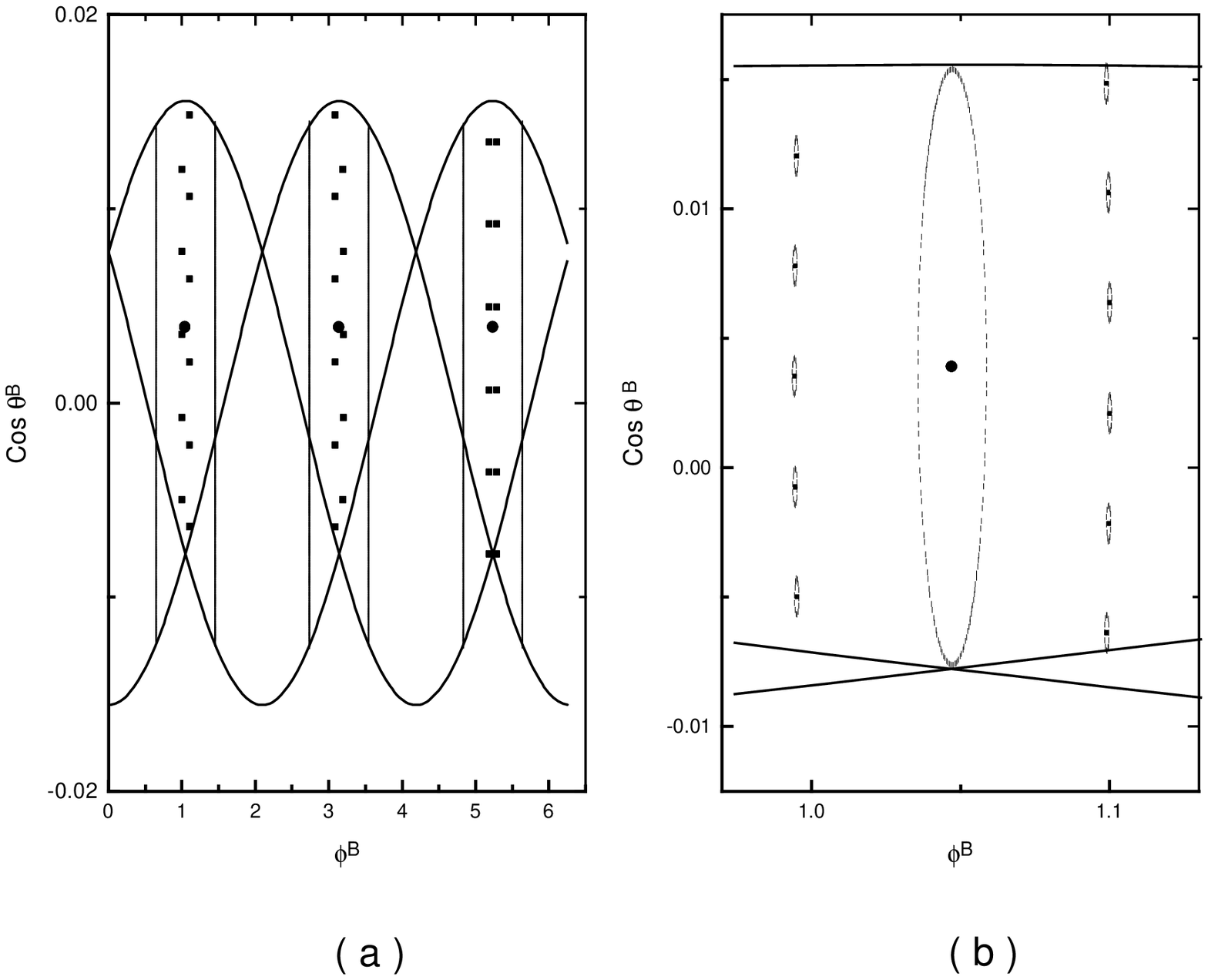,width=15.cm,height=20.cm,angle=0} } }
\vskip .5cm
\centerline{ \parbox{15cm} {{\normalsize {Fig.7 -- 
(a) Some tori in a planar triangle. Circles correspond to a rational torus with 
$n_p=3$, squares to $n_p=34$ and the straight line to an irrational torus. (b) For
a curved triangle with $\omega= {10001 \pi \over 30000 }$ open trajectories develope 
around the elliptic points.} }} }

\end{document}